\documentstyle[seceq,preprint,epsf]{jpsj}
\title
{Josephson Vortex States in Intermediate Fields}
\author{Ryusuke Ikeda}
\inst{Department of Physics, Kyoto University, Kyoto 606-8502, Japan}
\recdate{October 9, 2001}

\abst{
Motivated by recent resistance data in high $T_c$ superconductors under 
magnetic fields parallel to the CuO layers, we address two key issues on the 
Josephson-vortex phase diagram, an appearance of structural transitions on 
the observed first order transition (FOT) curve in intermediate fields 
and of a lower critical point of the FOT line. The structural transitions reflected on the thermal transition curve in YBCO are 
identified based on our mean field calculation, and it is argued that a lower 
critical point more 
easily occurs in the present ${\bf B} \perp c$ case compared 
with that in ${\bf B} \parallel c$ case. 
}

\kword{Vortex States}
\begin{document}
\sloppy
\maketitle

\section{Introduction}

Josephson-coupled layered superconductors under magnetic 
fields parallel to the layers (i.e., in ${\bf B} \perp c$) have at least two 
qualitatively different field ranges\cite{Th1,RI1}. In low enough fields defined by 
$p \equiv 2 \pi d^2 B \sqrt{\Gamma}/\phi_0 \ll 1$, the intrinsic 
pinning effect due to the layering is a weak perturbation and can be neglected 
for most purposes. Oppositely, in high enough fields $p \gg 1$, the layering 
effect is essential and strongly affects the $p$-dependence of 
fluctuation properties and of 
the transition or crossover lines. Here, $d$ is the layer spacing, 
$\sqrt{\Gamma} (> 1)$ the anisotropy of penetration depths, 
and $\phi_0$ the flux quantum. These two regimes are separated by an 
intermediate field range characterized by $p \sim 1$ in which, 
for instance, the transition or crossover temperatures 
defined in the disorder-free limit begin to lose their field 
dependences. Since this intermediate field range is 
reached by the applied fields of the order of tesla in BSCCO and underdoped 
YBCO, one needs to examine in details 
what are expected theoretically to occur in this 
field range of real systems including even some pinning disorder. 
Recent resistance data in high $T_c$ cuprates have suggested a first order 
behavior\cite{KK1,Zhukov,Naito,New} of the freezing (melting) transition 
in clean limit, the presence\cite{Zhukov,Naito,New} of 
its {\it lower} critical point (LCP) at least in YBCO (see below), 
and a reflection\cite{KK2,Zhukov} of field-induced 
structural transformations between different vortex solids. 

In refs.2 and 8, we have argued that the freezing to a solid of Josephson 
vortices in disorder-free limit in ${\bf B} \perp c$ is a first order transition (FOT) in all field ranges and that the FOT curve $T_m(B)$ in high fields $p > 1$ becomes insensitive to $p$ with 
increasing field. This is quite consistent with recent observations\cite{KK1,Zhukov,Naito,New}. We will focus below on the remaining two experimental findings 
mentioned above. First, to examine the 
reflection of structural transitions on the resistivity data, we need to clarify which types of vortex solids become ground states in clean limit in the relevant field range. As indicated in ref.2, in a narrow field 
range around a commensurate field where an intrinsically-pinned solid becomes 
the stable ground state due to a commensurability condition, the free energy in the solid phase is lowered reflecting a fluctuation-reduction due to the intrinsic pinning effect, and hence, the FOT temperature $T_m(B)$ is elevated. Since a unpinned solid or fractionally-pinned solid, i.e., a ground state with much weaker intrinsic pinning effect, can take the place of the pinned solid as the field slightly deviates from the commensurate field, an oscillating behavior of $T_m(B)$ may occur\cite{RI1} and, in fact, seems to have 
been found in resistivity data of underdoped YBCO\cite{Zhukov,Naito,New}. 
Although such an oscillating behavior was first speculated\cite{BN} for a putative vortex liquid-smectic transition, we do not believe\cite{RI1,RI2} the presence of vortex smectic phase at least in $p \geq 1$ where $T_m$ should begin to become field-independent. Recent resistivity data in {\it lower} fields in 
BSCCO \cite{KK2} have certainly shown a FOT between the vortex liquid and solid which, in contrast to the claim in ref.4, must {\it not} occur as the vortex smectic melting. This fact implies the absence of 
smectic phase even in $p < 1$. A key point is that the oscillating $T_m(B)$ 
curve was suggested in higher fields\cite{Zhukov,Naito,New} $p \geq 1$ 
where the intrinsic pinning effect is essential. Such a high field FOT line 
with oscillating behavior is 
consistent not with the vortex smectic-liquid transition scenario\cite{BN} 
but with the liquid-solid transition scenario with field ranges\cite{RI1}, 
in which the unpinned (floating) solid becomes the disorder-free ground state 
and hence, $T_m$ is lowered. 
According to the data\cite{Zhukov}, only a couple of field windows of pinned 
solids are observed. Since, in high enough fields ($p > 1.4$), 
field-induced structural transitions no longer 
occur\cite{RI1}, our theory may be able to clarify which vortex solids are 
realized in the observed oscillating melting behaviors. 
Calculated results on this issue will be given in $\S 2$. 

Another aspect relevant to the region $p \sim 1$, i.e., 
the presence of a LCP of the FOT curve, was suggested in resistivity data of YBCO with weaker fluctuation than of BSCCO: According to the underdoped YBCO data\cite{Zhukov,Naito,New}, the in-plane 
resistivity ($\rho_{ab}$) curves in lower fields suggest a continuous melting or glass transition, while $\rho_{ab}$ curves in higher fields have shown an abrupt decrease indicative of a FOT in the defect-free 
limit. Situation is similar to ${\bf B} \parallel c$ case\cite{RI3}, in which a LCP has been detected in overdoped YBCO but not in BSCCO with stronger fluctuation and larger anisotropy. Therefore, it is natural to guess that the second 
order $\rho_{ab}$-vanishing\cite{WKK} in tesla range of optimally-doped YBCO 
had been also a phenomenon 
in lower fields than LCP. Actually, {\it no} second order transition is 
suggested\cite{KK1,KK2} in the corresponding data in the low current limit, 
in both high and low fields, for 
clean BSCCO samples which have much stronger fluctuation and 
hence, a weaker pinning effect. Further, it has been quite recently 
clarified\cite{New} that the out-of-plane resistivity ($\rho_c$) in underdoped YBCO behaves in quite a similar manner to $\rho_{ab}$ and, 
for all fields, vanishes at essentially the same temperatures 
as those resulting from $\rho_{ab}$. 
Note that $\rho_c$ cannot vanish at a transition occuring without pinning disorder\cite{RI1}. Therefore, the observed continuous vanishing\cite{Zhukov,Naito,New,WKK} of resistivity in lower fields is likely to be due to a Josephson-vortex-glass (JG) transition\cite{RI2} occuring just {\it above} $T_m(B)$, 
while the almost discontinuous\cite{Zhukov,Naito,New} resistive vanishing, seen in higher fields where the intrinsic pinning effect is essential, reflects a 
JG transition induced\cite{RI4} by the FOT at $T_m(B)$. If 
this prediction is correct, an oscillating behavior suggested on both $T_m(B)$ curve and the glass transition line in high fields lying below $T_m(B)$ 
is not expected to appear on 
the continuous transition line in lower fields than LCP. 
However, a theoretical consideration is necessary in order to show that a 
situation in which the FOT is destroyed mainly 
by a LCP is not a rare event in real systems with random pinning effect under ${\bf B} 
\perp c$ and will be performed in $\S 3$. 
Finally, in $\S 4$, our theoretical results are compared with available 
resistivity data in ${\bf B} \perp c$. 
 
%However, no oscillating behavior has been seen on the melting line in BSCCO, al%though an evidence of a couple of field-induced structural transitions between %vortex solids has been found in the data under tilted magnetic fields. This 
%difference between YBCO and BSCCO may be due to a large difference in the 
%fluctuation strength. 

\section{Sequence of Structural Transitions in $p \sim 1$} 

In this section, we reexamine the field-induced sequence of structural 
transitions between Josephson vortex solids, particularly in the intermediate 
fields around $p \simeq 1$. Structural transitions in this region were 
considered only briefly in the previous works\cite{RI1} (see Fig.7 (a) in ref.2) and will be compared in details on this occasion with available 
data\cite{Zhukov,Naito}. In addition, a recent study\cite{Argen} using the 
London model with uniaxially periodic potential has suggeted 
an importance of rotated (pinned) solids in ${\bf B} \perp c$ 
in which all vortices are pinned between two neighboring layers. 
Here, we primarily examine energies and stabilities of 
rotated solids possible around $p \sim 1$. In $\S 4$, it will be pointed out that a couple of rotated solids are good candidates of ground states reflected in the observed oscillating melting (and glass) curves. 

Following ref.2, we apply the LLL approximation of order parameter $\psi$, 
defined\cite{RI1} in continuous $x$-$y$ space, to the Lawrence-Doniach model  
$${\cal H}_{\rm LD} = \int dy \int dx \sum_{m=-\infty}^{\infty} \exp\biggl({\rm i}{{{2 \pi} m x} \over d} \biggr) \biggl[ \varepsilon_0|\psi(x,y)|^2 + \xi_0^2 |(-{\rm i} \partial_y + r_B^{-2} x) \psi(x,y) \, )|^2 $$ $$+ \Gamma^{-1} d^{-2} \xi_0^2 |\psi(x,y) - \psi(x+d, y)|^2 + {b \over 2} |\psi(x,y)|^4 \biggr] \eqno(2.1)$$
rewritten\cite{RI1} in terms of the Poisson-summation formula, where $\xi_0$ is the in-plane coherence length, $r_B = \sqrt{\phi_0/(2 \pi B)}$, $b >0$ a constant, and $\varepsilon_0 \simeq (T-T_{c0})/T_{c0}$. Spatial variations parallel to ${\bf B}$ were neglected by focusing on the mean field solutions. Further, in the $x$-$y$ plane perpendicular to ${\bf B}$, the out-of-layer direction and the in-layer 
direction are denoted as $x$ and $y$-directions, respectively, and the type II limit with no variation of flux density was assumed. In this representation, the "core" of a Josephson vortex is {\it expressed} as that of the ordinary singular vortex. Since the thickness of each superconducting layer is infinitesimal in the Lawrence-Doniach model, all vortex "cores" are situated between the 
superconducting layers. In a single vortex picture, each vortex is expected to 
sit in the middle of an inter-layer spacing. However, the type II limit assumed here cannot reduce to such a single vortex picture because the interaction-range between the field-induced vortices is infinite in type II limit, and hence, 
the vortices in type II limit may form a solid structure favored by the 
long-ranged interaction between them 
without sitting in the middle of inter-layer 
spacings. Actually, it is found\cite{RI1} that fractional pinned solids, in 
which a fraction of vortices are not located in the middle of inter-layer spacings, can become ground states in clean limit at fields determined by their 
commensurability conditions. 

 A periodic solution including 
rotated lattices is expressed, within LLL in this representation, in the form 
$$\psi(x,y) = \sqrt{\Gamma^{1/4} {{k r_B} \over {\pi^{1/2}}}} \sum_m \exp({\rm i}kmy - {{\sqrt{\Gamma}} \over {2 r_B^2}}(x+k m r_B^2)^2 + {{{\rm i}\pi} \over R}m^2), \eqno(2.2)$$
where $R$ is a rational number. Although several rotated (pinned) solids 
in which $R$ takes integer values larger than 2 were studied in ref.2 (see 
the sentences below eq.(2.4) there), they were not identified with ground 
states, possibly because of too low field values. As shown below, 
rotated structures with a fractional $R$ can have energy remarkably lowered 
due to the vortex confinement induced by the layering and hence, become the 
ground states in the intermediate fields $p \sim 1$. Although the rotated version of the fractional pinned solids can also be created by choosing a fractional $R$, most of them seem to be fragile in real systems with 
disorder\cite{RI1} (see $\S 4$). Here we will focus on strongly 
(intrinsically-) pinned solids called in 
ref.2 integer pinned solids, in which each vortex is located 
in the middle of an 
inter-layer spacing. An integer pinned solid is defined by $kr_B^2 = wd$ with 
a positive integer $w$ and has $(w-1)d$ vacant inter-layer spacings between 
two neighboring occupied inter-layer 
spacings. Namely, in a $w=1$ solid all inter-layer spacings are occupied by 
vortices. Further, we will not consider here integer solids 
with $w \geq 2$ but will focus on rotated $w=1$ solids for a reason to be 
clarified in $\S 4$. 

Based on eq.(2.2), the rotated $w=1$ solids are created as follows. The 
nonrotated $w=1$ solid is represented in terms of the parameters, $R=2$ and 
$k r_B^2=d$, and have the basis vectors ${\bf a}^{(1)} = (2 \pi/k){\hat y}$ and ${\bf a}^{(2)} = (\pi/k){\hat y} - w d{\hat x}$ (see Fig.1 (a)). Now, let us 
rotate these basis vectors 
counter-clockwise by an angle $\theta$, ${\bf a}^{(j)} 
\to {\bf a}^{(j)}(\theta)$. It is easy to find that the condition under which 
the resulting rotated solid is intrinsically pinned by the layers 
is $Ma^{(1)}_x(\theta)=Na^{(2)}_x(\theta) = MNd$, where $M$ and $N$ are 
positive integers. 
The resulting structure is represented 
within LLL in terms of eq.(2.2) by setting $kr_B^2=d$ and 
$$R \to R_{MN} = {{2 (M^2+N^2-MN)} \over {2M-N}}, \eqno(2.3)$$  
and will be called below $(M$, $N)$ state. The commensurability condition 
of this structure is 
$$p = {{d^2 \sqrt{\Gamma}} \over {r_B^2}} = {{\sqrt{3} \pi} \over {M^2+N^2-MN}}, \eqno(2.4)$$ 
which means that the state $(N$, $M)$ is denenerate in energy 
with the $(M$, $N)$ state. 
Actually, it is seen that they can be transformed to each other 
by reversing the field direction ${\bf B} \to -{\bf B}$. 
According to the expressions mentioned above, the rotated 
states with $N=1$ or $M=1$ belong to $w=1$ 
solids with no vacant inter-layer spacings. 
Since a more energy gain due to the 
intrinsic pinning is expected to be maximized in 
a $w=1$ state, we will primarily 
consider states with $N$ or $M=1$. Actually, for instance, the $(2$, $3)$ state is equivalent to the $(3$, $1)$ one. Further, as shown previously\cite{RI1}, the commensurability conditions for nonrotated integer solids with $(w-1)d$ vacant spacings in the $x$-direction are $p=\sqrt{3} \pi/w^2$ 
and $\pi/(\sqrt{3} w^2)$. From this and eq.(2.4), the $(M$, $M)$ 
and $(2M$, $M)$ states are found to be merely alternative representations 
of nonrotated solids. 
 
 The rotated $w=1$ solid to be realized in a highest field is the $(1$, $3)$ state, illustrated in Fig.1 (b), or its reversal one $(3$, $1)$ (Fig.1 (c)), and their commensurate field is given by 
 $p_{13}=\sqrt{3}\pi/7 \simeq 0.77$. The rotated 
 state realizable in the next lowest 
 field, with decreasing field, is the $(1$, $4)$ or $(4$, $1)$ state which becomes commensurate when $p_{14}=\sqrt{3} \pi/13 \simeq 0.42$. Since the layering is not effective with decreasing field, other rotated 
 states occuring by lowering field 
 further will not be considered. 

 Keeping these commensurate fields for various structures in mind, the sequence of ground states changing with increasing $p$ can be determined\cite{RI1} by computing the value of Abrikosov factor $\beta_{\rm A}$ of each structure for various $p$-values. The resulting sequence of ground states is given by the table of Fig.2, where we take account only of the $w=$ integer nonrotated solids and $(M$,$1)$ rotated solids as candidates. In the $p$-ranges shown in blank spaces of the upper row of Fig.2, fractional pinned solids or the unpinned (floating) solid become ground states. It is found that the $w > 1$ nonrotated integer solids become ground states just at their commensurate field values, while a $(M$, $1)$ state, particularly the $(3$, $1)$ state, is the lowest in energy over a {\it broad} range 
 including its commensurate field. In the case of $(3$, $1)$, $\beta_{\rm A} < 1.15959$ (the value for the floating solid) in the range $0.725 < p < 0.96$ including its commensurate field $p=0.77$, and $\beta_{\rm A}$ reaches its lowest value 1.157 near $p=0.8$. Further, we found that the $(4$, $1)$ state becomes the ground state in $0.416 < p < 0.421$. This result that, at least in intermediate fields specified by $p$-values of order unity, a rotated $w=1$ solid is minimized over a broad range including its commensurate field value is a reflection of the intrinsic pinning effect enhanced by the 
 absence of vacant inter-layer spacings. Thus, it may be possible that, if the model (2.1) is analyzed beyond the LLL approximation, a rotated $w=1$ solid becomes the ground state at a commensurate field of a $w>1$ nonrotated solid. 
 Actually, we have found in ref.2 that the $w=1$ nonrotated solid is lower in 
 energy than the $w=2$ one when $p=\sqrt{3} \pi/2^2$. 
 
   It should be noted that the period in the $x$ direction of the $(M$, $1)$ state is $2(1+M^2-M)d$, while the corresponding period of the $w=2$ nonrotated solid is $4d$. Hence, to find the $(M$, $1)$ solid in a numerical simulation in fields near its commensurate field, setting the system size in the $x$ direction to a multiple of $2(1+M^2-M)d$ is necessary. Otherwise, a numerical simulatioin would create a wrong ground state at lower temperatures.  

Before closing this section, 
it is necessary to clarify whether a superposition of the two states, $(M$, $1)$ and $(1$, $M)$, results 
in a lowering of energy, since they are degenerate 
in energy with each other. In fact, the authors of ref.13 have argued that a mixing of the two pure states will be a representation of a waving vortex 
solid structure. On the other hand, the present author has given an evidence of such a waving solid in a limited field range ($1.26 < p < 1.4$) by 
showing\cite{RI1} that this structure is created due to a shear instability of the $w=1$ nonrotated solid. 
To examine whether the former 
mechanism\cite{Argen} is possible, we have computed the energies of various superpositions of the $(1$,$3)$ and $(3$, $1)$ states in the range $0.72 < p 
< 0.93$ (see Fig.2) and found that such a mixing always raises the energy. 
Namely, in fields around the commensurate field, either the $(1$,$M)$ or $(M$, $1)$ states is spontaneously realized as the ground state. Actually, we have checked that the shear modulus of the $(1$, $3)$ state is positive in the above-mentioned field range, guaranteeing the stability of pure rotated states. It is physically natural that a mixing of the two pure rotated states does not result in 
a waving ground state solid: The picture on the waving solid argued in ref.13 requires that, when moving along the $x$ direction ($\perp$ layers), the $(1$, $M)$ and $(M$, $1)$ states alternate with domain walls intervening between them. Due to such domain walls with positive energy, a pure rotated solid should have a lower energy compared with any mixing of two equivalent pure rotated solids. 

\section{Presence of a Lower Critical Point} 

In ref.8, we have studied the JG 
transition in ${\bf B} \perp c$ in higher fields ($p \gg 1$), where the FOT curve $T_m$ in {\it disorder-free} case is insensitive to $p$, and the glass transition temperature 
$T_{\rm JG}$ is lower than $T_m(p)$. However, a remarkable deviation of $T_{\rm JG}(p)$ from $T_m(p)$ to lower temperatures has been seen only in dirtier BSCCO under higher fields\cite{Naka} where the FOT in ${\bf B} 
\perp c$ was not detected. In underdoped YBCO with much smaller anisotropy than in BSCCO, the $T_{\rm JG}(p)$-line in $p \gg 1$ does not deviate remarkably from $T_m(p)$ 
particularly in cleaner samples\cite{RI2} (see also $\S 4$). 

Rather, a continuous vanishing of in-plane resistivity $\rho_{ab}$ in YBCO has been often observed in lower fields where $T_m(B)$ is sensitive to $B$. 
Taking account of the corresponding $\rho_c$-vanishing\cite{New}, this continuous transition is, just like that explained in ${\bf B} \parallel c$ case\cite{RI3}, likely to be the JG transition\cite{RI2} occurring {\it below} a LCP. Further, since no discontinuous resistive vanishing occurs in the same sample\cite{Naito} in ${\bf B} \parallel c$, we feel that the FOT in ${\bf B} \perp c$ case tends to be destroyed, in contrast to ${\bf B} \parallel c$ case, primarily 
from the 
lower field side by creating a LCP. 
To show that this is indeed so, we consider an extrapolation to higher fields of a JG curve computed in $p \ll 1$ which, just as the thin solid curve shows 
in Fig.3, lies in $p < p^*$ {\it above} $T_m(B)$ due to point disorder with moderate 
strength. Based on the familiar\cite{RI4} results in ${\bf B} \parallel c$ case, one might expect the $T_{\rm JG}(B)$ line to have a functional 
form similar to $T_m(B)$ (described as the chain curve in Fig.3) even in ${\bf B} \perp c$ case and hence, to become insensitive to $B$ in higher fields. If so, the FOT would not occur at any field because the resulting JG transition curve would always lie above $T_m(B)$, and the fact, that $T_{\rm JG} > T_m$ in lower fields, would imply that the FOT be destroyed mainly through a reduction of an upper critical point. As is seen below, however, 
this picture does not apply to the ${\bf B} 
\perp c$ case. 

Upon cooling from higher temperatures, the JG transition is signaled by a divergence of the glass susceptibility\cite{FFH,RI2} $\chi_{\rm G}$ calculated in terms of the model (2.1), and the diagrammatic form of $\chi_{\rm G}$ is typically a series of ladder diagrams composed of bare disorder (random pinning) lines. As shown in ${\bf B} \parallel c$ case\cite{RI4}, the vertex correction to each bare disorder line arising from a precursor of vortex solidification enhances the disorder strength and $T_{\rm JG}$, while the vertex correction originating from a high temperature fluctuation effect unrelated to the vortex solidification reduces $T_{\rm JG}$. The former effect in high field limit for ${\bf B} \perp c$ case was studied in ref.8. Here, we consider, by assuming $T_{\rm JG} > T_m$ in lower fields, just the latter effect which can be represented by the familiar diagrams of RPA type. In low fields where the layer structure is negligible, the $\chi_{\rm G}$-expression is essentially the same as that for the isotropic 3D GL model and, in LLL approximation, gives a $T_{\rm JG}$-curve yielding the LLL scaling\cite{RI5} and may lie {\it above} $T_m(B)$ for a moderate disorder strength. Since it is difficult to directly perform the corresponding calculation in the intermediate fields, let us consider here the high field limit ($p \gg 1$), in contrast to in ref.8, by neglecting the vertex correction related to the vortex solidification effect. As is intuitively clear, the model (2.1) loses all field dependences in this limit and takes the same form as the GL model in 2D and {\it zero} field except for the shift $T_{c0} - T_{c2}$ of the position of its mean field transition (see $\S 4$). On the other hand, it is well known\cite{RI3} through formally the same calculation as that perfomed in the context of spin-glass\cite{Sh} that a superconducting-glass transition in homogeneous (i.e., nongranular) materials under zero field does not occur even at the mean field level (i.e., irrespective of dimensionality). Actually, since, in the irreducible vertex in the ladder representing $\chi_{\rm G}$, the fluctuation vertex correction to the disorder 
strength 
outweighs the temperature dependence accompanying the bare disorder strength, the irreducible vertex itself {\it decreases} on cooling and consequently, $\chi_{\rm G}$ cannot diverge at any temperature even if the renormalization of glass fluctuations, leading to a dimensionality dependence, is neglected. This result $T_{\rm JG}(p \to \infty) \to 0$ obtained without effects of vortex solidification, when combined with the result on $T_{\rm JG}$ in ref.8 described in Fig.3 as the thick solid curve, implies that, as described by the solid curves in Fig.3, any $T_{\rm JG}$-line lying above $T_m$ in lower fields {\it must} cross $T_m(B)$-curve at some field and lie at lower 
temperatures than the disorder-free $T_m(B)$-curve (the chain curve in Fig.3) in higher fields. Clearly, a point essential to reach this conclusion is that, in contrast to in ${\bf B} \parallel c$ case, the $T_m(B)$-curve becomes insensitive to $p$ in $p > 1$, i.e., that $T_m$ drastically changes its $p$-dependence in an intermediate field $p \sim 1$, while $T_{\rm JG}$ has no such a property. Therefore, a LCP in ${\bf B} \perp c$ case is expected to, in many cases, be created in the range $p \sim 1$ so that the FOT in $p > 1$ with strong intrinsic pinning effect remains.   
 
\section{Discussion}

Based on our results in the preceding sections, let us discuss here the doping dependences of resistivity data in ref.4 and the resulting phase diagram in ${\bf B} \perp c$. As mentioned in $\S 1$ (see also the 
comment\cite{Zhukov} on ref.4), the oscillating behavior on the vertical portion of the FOT line is a clear evidence that the transition corresponds not to the putative liquid to smectic transition but to a liquid to solid (strictly speaking, liquid to slush\cite{RI4,RI3}) transition argued in ref.2. In high fields $p \gg 1$, this liquid to solid (slush) transition line $T_m(p)$ was 
denoted in ref.8 as $T_{sc}$ and, according to $\S 5$ in ref.2, is given by 
$$T_m(p \gg 1) \simeq T_{c0} ( 1 - {{2 \xi_0^2} \over {d^2}} \Gamma^{-1}) (1+ c_\infty \, \varepsilon_G^{(2)} )^{-1} \eqno(4.1)$$
(see the paragraph including eq.(5.1) in ref.2), where $\varepsilon_G^{(2)} = 16 \pi^2 k_{\rm B} T_{c0} \lambda^2(0)/(\phi_0^2 d)$ is 2D Ginzburg number, 
$\lambda(0)$ the in-plane penetration depth, $c_{\infty} (>0)$ is a constant of order unity, and the combination $T_{c2} \equiv T_{c0}(1 - 2 \Gamma^{-1}(\xi_0/d)^2)$ in eq.(4.1) expresses the mean field transition temperature in $p \to \infty$ limit\cite{RI1}. Now, let us compare eq.(4.1) with Fig.4 in ref.4, which shows that the transition temperature vertical (i.e., independent of $B$) in higher fields {\it increases} with underdoping. Since both $\Gamma$ and $\lambda(0)$ increase with underdoping, while a doping dependence\cite{Zhukov} of $T_{c0}$ is negligible in the doping range of 
samples in ref.4, the doping dependence of the vertical transition line is primarily due to the $\Gamma$-dependence of eq.(4.1), i.e., of $T_{c2}$. Namely, the doping dependence of $\varepsilon_G^{(2)}$ may be neglected compared with that of $\Gamma$. We emphasize that, in the phase-only approximation of the Lawrence-Doniach model, $T_{c2}$ is obtained as $T_{c0}$, 
and hence that the above-mentioned 
doping dependence of $T_m(p > 1)$ cannot be explained in terms of phase-only 
models\cite{Th1} such as the XY model. 

Next, let us consider at what field $T_m(B)$ begins to become insensitive 
to $B$. To clarify this, it will be sufficient to see at what field, called 
$B_{\rm X}$ hereafter, 
the portions of $T_m(B)$ in higher ($p \gg 1$) and lower ($p \ll 1$) 
fields meet when they are interpolated with each other. Since eq.(4.1) is independent\cite{RI2} of an approximation method, while the portion in $p \ll 1$ where the layering is a secondary effect depends highly on 
the approximation methods, 
we need to consider under what condition the LLL approximation for the $p < 1$ 
region becomes inapplicable. Estimating a crossover field $B_{cr}^{(3)}$ between the LLL and the London behaviors in ${\bf B} \parallel c$ case was performed in ref.16 (see eq.(4.1) in ref.16). Within the anisotropic 3D GL 
model\cite{RI6}, a crossover field in ${\bf B} \perp c$ corresponding to this 
is easily found, and the LLL approach is semi-quantitatively valid when 
$$B > \sqrt{\Gamma} B_{cr}^{(3)}. \eqno(4.2)$$
In ref.16, $B_{cr}^{(3)}$ was estimated to be 1.7 tesla for the 
optimally-doped YBCO. Using this and based on eq.(4.2), 
the r.h.s. of eq.(4.2) may be, due to its $\Gamma$-dependence, 
beyond $50$ (T) for underdoped YBCO even if the doping dependence of $\lambda(0)$ is neglected. Through this fact, the $T_m(B)$-curves in $p < 1$ for underdoped YBCO will be assumed to be well described in the London limit of anisotropic 3D GL model. The corresponding $T_m(B)$ is nothing but $T_m^{(c)}(p)$ curve given in ref.8 and has the form $T_m^{(c)} \simeq T_{c2}(p) (1 - \varepsilon_G^{(2)} \sqrt{p})$. By comparing this with eq.(4.1), $B_{\rm X}$ appropriate to 
underdoped YBCO is defined assuming $p$ to be a constant, i.e., $B_{\rm X} \propto \Gamma^{-1/2}$, where $d$ was assumed to be doping-independent. By analyzing Fig.4 in ref.4 in terms of this $B_{\rm X}$ and eq.(4.1), we find that, for each sample in ref.4, 
$\sqrt{\Gamma} \simeq$ 32 (T750), 14 (T700), and 8 (T650), by assuming other parameters to be doping-independent. Consistently with these $\Gamma$ values, we find $c_\infty \varepsilon_G^{(2)} \simeq 0.043$. It means that, when taking a reasonable value $\lambda(0) \simeq 10^3$ (A), the number $c_\infty$ is 
about 2.0 which coincides with the number\cite{HN} giving the Kosterlitz-Thouless transition point in 2D films in zero field. This is reasonable 
because the present problem would reduce to the 2D system in zero field {\it if the shift $T_{c0}-T_{c2}$ of the mean field boundary be absent}. 

We next try to describe the oscillating $T_m(B)$ of the sample T750 in ref.4 by ascribing its origins to a couple of pinned ground states. Figure 2 in ref.4 shows that the oscillating $T_m$ and $T_{\rm JG}$ lines suggested from the resistivity data include three stability regions of pinned solid states in fields below 14 (T). The FOT at $T_m(B)$ in clean limit seems to be reflected in the data above 4 (T) {\it and below} 12 (T). Namely, both upper and lower critical points should be present there: 
We note that the resistive vanishing at 14(T) in Fig.1 of ref.4 is clearly continuous and that the presence of a upper critical point below 14 (T) 
is expected. 
Of course, the feature that the transition curve is vertical ($B$-independent) above 4 (T) originates from the intrinsic pinning effect due to the layering. One of remarkable features is that the middle window, suggestive of a pinned ground state structure, around 6 (T) is remarkably broad. First, the large $\Gamma$ ($\simeq 1 \times 10^3$) value of the sample mentioned above implies that the $p \simeq 1$-value for this sample corresponds to about 10 (T). Since, according to the present Fig.2, no structural transition will occur in $p > 1.26$ of real systems\cite{com01}, the suggested field-induced transition near 10 (T) will imply an onset of the $w=1$ waving solid lying near $p=1.26$. 
Then, turning to lower fields, it is quantitatively reasonable to assume that the wider window, suggestive of a stronger intrinsic pinning, around $B \sim $ 6(T) will correspond to the window in Fig.2 in which the $(1$,$3)$ or $(3$, $1)$ state is stable. As mentioned in $\S 2$, the rotated $w=1$ solids such as a $(M$, $1)$ state are favored, even if the field deviates from their commensurate field, by the confinement effects of the layering and may have their wide widths in which these solids have the lowest energy. This tendency is lost with increasing $M$. Hence, a width of a stability region of a $(M$, $1)$ state relevant to lower fields should be narrower. If the $w>1$ solids are not realized, the third pinned region near 3.3(T) suggested in Fig.1 of ref.4 will be a reflection of the (1,4) (or (4,1) ) state. Although the commensurate field $p=0.42$ of the (4,1) state suggests that, in the data of ref.4, this state should be realized in slightly lower fields, and the $w=2$ nonrotated solid is more reasonable in view only of commensurate field values, a rotated solid may be minimized, due to the vortex confinement effect, at a commensurate field of a $w>1$ nonrotated solid (see $\S 2$). In fact, although a $w=3$ nonrotated solid has its commensurate field $p=0.59$ and hence, is expected to be realized near 5(T) in Fig.2 of ref.4, there is no sign of realization of the $w=3$ state in data of ref.4, possibly 
reflecting its disappearance\cite{RI1} due to the random disorder. Through 
this fact, there 
will be no reason why the $w=2$ state should be realized in the 
same {\it real} sample. Therefore. we expect that the third pinned state near 3.3 (T) will correspond to the (4,1) (or (1,4)) 
state belonging to the $w=1$ states. 

Here, we comment on the positions of $T_{\rm JG}(B)$ {\it above} an upper critical point (UCP) of underdoped YBCO according to the result in ref.8. Broadly speaking, $T_{\rm JG}(B)$ above UCP is less steep with increasing $B$, and the JG-transition curve derived in ref.8 
$$B_{\rm JG}(T) \simeq B_0 {{T_c^*(\Delta) - T} \over T}, \eqno(4.3)$$
becomes applicable with increasing $B$, 
where $B_0 = \phi_0/(2 \pi d^2 \sqrt{\Gamma} \varepsilon_G^{(2)})$, and $\Delta$ denotes the random pinning strength. Since, assuming that only $\Gamma$ is doping-dependent and that 
$\varepsilon_G^{(2)} \simeq 0.02$ and $d \simeq 10 (A)$, $B_0$ for YBCO is estimated to be $10^4/\sqrt{\Gamma}$ tesla, the $B_{\rm JG}(T)$ curve for underdoped YBCO in ref.4 with $\Gamma < 10^3$ is found to be more steep than the corresponding $H_{c2}(T)$-line in ${\bf B} \parallel c$ case. This is in contrast to that in BSCCO which was primarily considered in ref.8 and found to be much flatter in the $B$-$T$ diagram. 

It will be necessary to explain why an oscillating behavior of $T_m(B)$ and its LCP were not observed in the corresponding data\cite{KK1,KK2} of BSCCO. The absence of LCP in BSCCO under ${\bf B} \perp c$ does not contradict its presence in YBCO. As in ${\bf B} \parallel c$ case\cite{RI3}, its origin can naturally be ascribed to a stronger fluctuation and a much larger anisotropy in BSCCO both of which reduces the LCP. Further, since the FOT in ${\bf B} \parallel c$ is present in the sample in ref.7 in contrast to the YBCO sample in 
ref.5, a weaker disorder is anticipated for the sample\cite{KK2}, which also results in a reduction of LCP. Rather, it should be noted that structural transitions of Josephson vortices have not been reflected at least in the vertical portion\cite{KK1} of the resulting FOT curve. Instead, a couple of structural transition fields suggested from the data\cite{KK2} in tilted fields were found to remarkably decrease with increasing temperature. By examining 
the BSCCO data\cite{KK1,KK2} altogether, the structural transition fields near $T_m(B)$ likely lie below $B_{\rm X}$ (see a preceding paragraph), i.e., in lower fields than the vertical portion of $T_m(B)$. This observation cannot be explained within the mean field approximation and suggests a strong $T$-dependence of the structure transition fields. This strong $T$-dependence may be due to an effective reduction of anisotropy occuring from the fact that the thermal fluctuation in BSCCO is not negligible even 
below $T_m(B)$. To resolve this issue, a quantitative study of fluctuation effects in the solid phases is needed and left to a future work.  @

%\leftline{Acknowledgement}

\vfil\eject

\leftline{Figure Captions} 

Fig.1 

(a) Basis vectors ${\bf a}^{(1)}$ and ${\bf a}^{(2)}$ in the $w=2$ nonrotated solid described in the $x$-$y$ plane, where the horizontal lines denote the superconducting layers, and the open circles indicate the "cores" of Josephson vortices. (b) Vortex configurations in the $w=1$ rotated solid $(3$,$1)$, and (c) the one in the $(1$,$3)$ state. 

\vspace{15mm}

Fig.2  

Table expressing the sequence of ground states at low enough temperatures. Here, the states denoted as $w=$ integer imply the nonrotated integer pinned solids, and $(M$,$1)$ states are rotated $w=1$ solids. The blank regions on the upper row, such as the one including the $p=1$ value, imply the field ranges in which 
the unpinned or fractionally pinned solids become ground states. Note that the structures expected to occur in $p < 0.4$ are not shown. 

\vspace{5mm}

Fig.3

Schematic phase diagram in ${\bf B} \perp c$. Thin solid curve implies the $T_{\rm JG}(B)$ curve obtained by assuming $T_{\rm JG}(B) > T_m(B)$ (the chain curve), and hence, only its portion in $p < p^*$ (i.e., below the LCP) plays the role of true $T_{\rm JG}(B)$. The true $T_{\rm JG}(B)$ curve in $p > p^*$, examined in ref.8, corresponds to the thick solid curve. The thermal FOT cannot\cite{RI3} occur on the portion in $p < p^*$ of the chain curve. 

\begin{figure}[t]
\begin{center}
\leavevmode
\epsfysize=15cm
\epsfbox{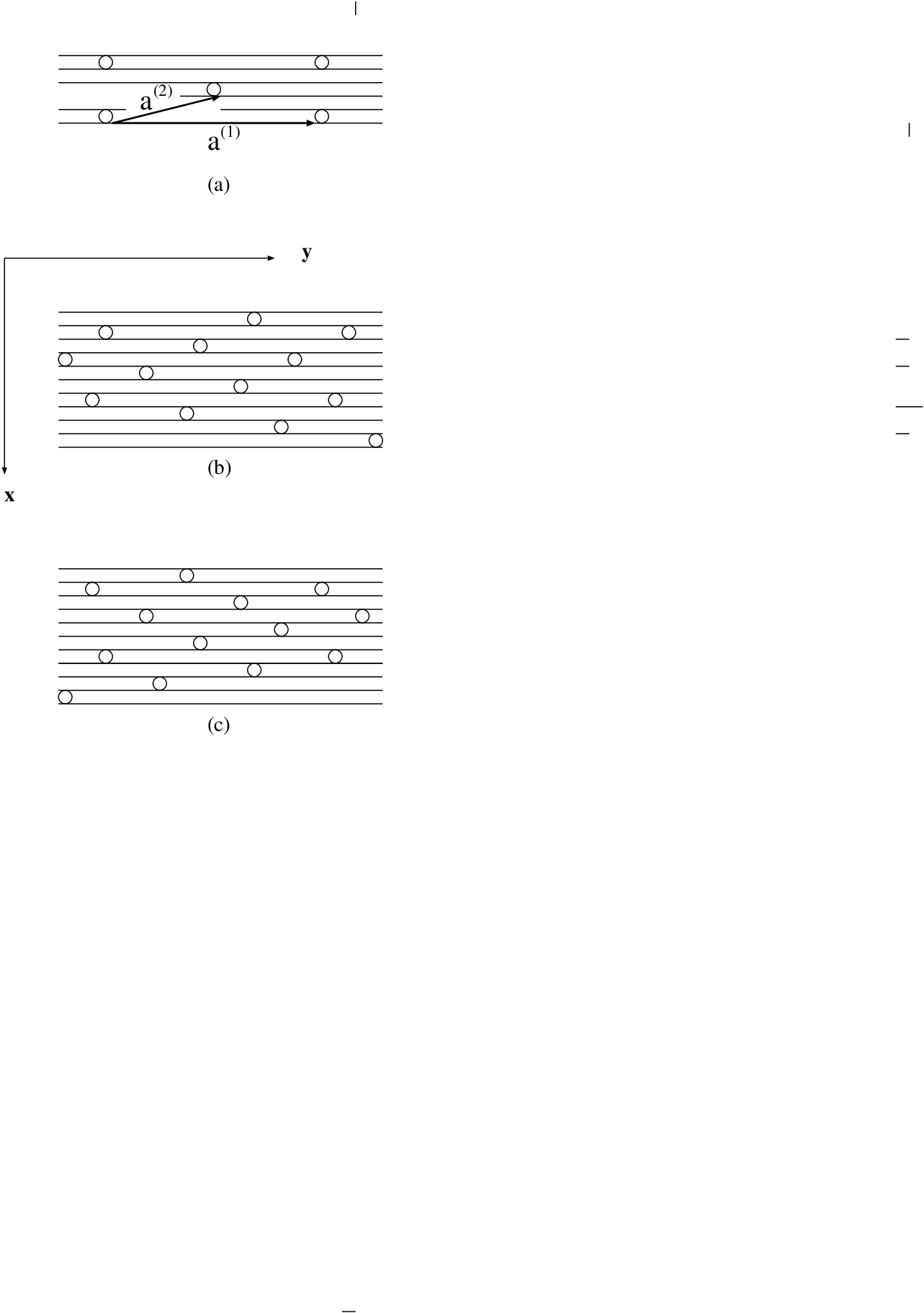}
\end{center}
\end{figure}

\begin{figure}[t]
\begin{center}
\leavevmode
\epsfysize=17cm
\epsfbox{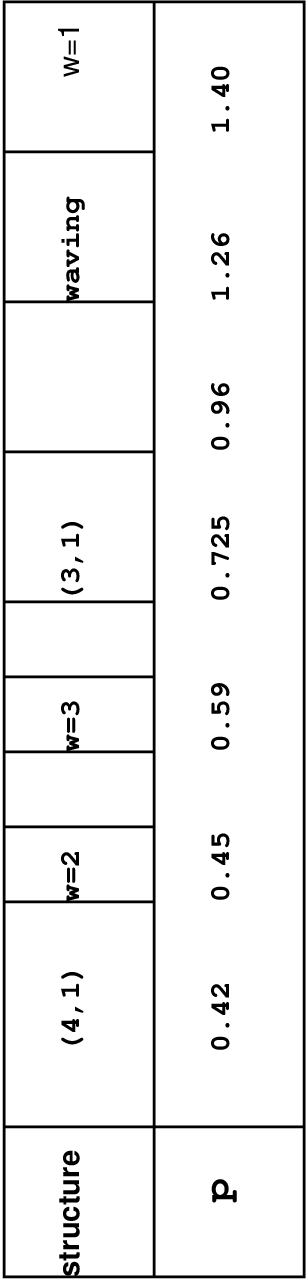}
\end{center}
\end{figure}

\begin{figure}[t]
\begin{center}
\leavevmode
\epsfysize=13cm
\epsfbox{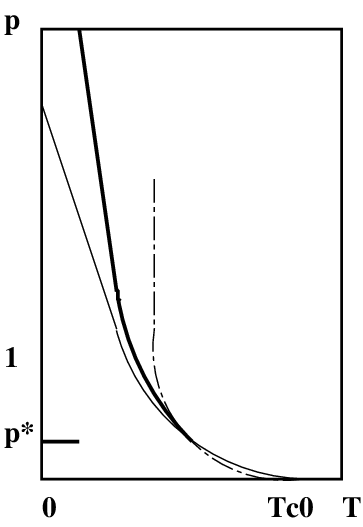}
\end{center}
\end{figure}


\begin{thebibliography}{99}
\bibitem{com01} As far as the thermal fluctuation is neglected, the energy of 
the $w=1$ waving solid per unit cell is almost the same as that of the w=1 
regular pinned solid, and the structural transition between them will be of a very weak first order which will be destroyed easily by point disorder in real 
systems.
\end{thebibliography}
\end{document}